\documentclass{ICSnewsletter}

\usepackage{lipsum}
\usepackage{cite}
\usepackage{amsmath,amssymb,amsfonts,bm,upgreek}
\usepackage{tabularx,booktabs}
\usepackage{algorithm,algpseudocode}
\usepackage{graphicx}
\usepackage{textcomp}
\usepackage{xcolor}
\def\BibTeX{{\rm B\kern-.05em{\sc i\kern-.025em b}\kern-.08em
    T\kern-.1667em\lower.7ex\hbox{E}\kern-.125emX}}
\usepackage{fitmc}
\newcommand{\trans}{\!\top} % transposed nicer
\newcommand{\dndt}{\frac{\mathrm d}{\mathrm dt}}

\DeclareMathOperator{\divergence}{div}
 
\DeclareMathOperator{\curl}{curl} \DeclareMathOperator{\grad}{grad}

\begin{document}

\makeatletter
\newcommand{\linebreakand}{%
  \end{@IEEEauthorhalign}
  \hfill\mbox{}\par
  \mbox{}\hfill\begin{@IEEEauthorhalign}
}
\makeatother

\title{Electromagnetic Quasistatic Field Formulations of Darwin Type
\\
\footnotesize
%\thanks{Corresponding author: M.~Clemens (email: clemens@uni-wuppertal.de)\\This work was %supported in parts by the Deutsche Forschungsgemeinschaft (DFG) under grant no. %CL143/11-2.}
~\\
\noindent Markus Clemens$^1$, Marvin-Lucas Henkel$^1$, Fotios Kasolis$^1$, Michael G\"unther$^2$,  Herbert De Gersem$^3$, Sebastian Schöps$^3$
~\\[1em]
\begin{minipage}[t]{0.325\textwidth}%
\normalfont
{$^1$}University of Wuppertal,\\
Chair of Electromagnetic Theory,\\ 
Rainer-Gruenter-Str. 21, 42119 Wuppertal, Germany\\
clemens@uni-wuppertal.de\\
www.tet.uni-wuppertal.de
\end{minipage}%
\begin{minipage}[t]{0.34\textwidth}
\normalfont
{$^2$}University of Wuppertal,\\
Chair of Numerical Analysis and Applied Mathematics\\ 
Gau{\ss}straße 20, 42119 Wuppertal, Germany\\
guenther@math.uni-wuppertal.de\\
www.imacm.uni-wuppertal.de
\end{minipage}%
\begin{minipage}[t]{0.35\textwidth}
\normalfont
{$^3$}Technical University of Darmstadt,\\ 
Inst. for Accelerator Science and Electromagnetic Fields\\
Schlo{\ss}gartenstr. 8, 64293 Darmstadt, Germany\\
schoeps@temf.tu-darmstadt.de,\\ degersem@temf.tu-darmstadt.de\\
www.temf.de
\end{minipage}%
~\\[1em]
\it This paper is dedicated to the lifetime work on electromagnetic field formulations of Professor Thomas Weiland on the occasion of his 70$^{\text{th}}$ birthday.
}

\maketitle

\begin{abstract}
Electromagnetic quasistatic (EMQS) fields, where radiation effects are neglected, while Ohmic losses and electric and magnetic field energies are considered, can be modeled using Darwin-type field models as an approximation to the full Maxwell equations. Commonly formulated in terms of magnetic vector and electric scalar potentials, these EMQS formulations are not gauge invariant. Several EMQS formulations resulting from different gauge equations are considered and analyzed in terms of their structural properties and their modeling capabilities and limitations. Associated discrete field formulations in the context of the Maxwell-grid equations of the Finite Integration Technique are considered in frequency and time domain and are studied with respect to their algebraic properties. A comparison of numerical simulation results w.r.t. reference solutions obtained with established formulations for the full Maxwell equations are presented. 
\end{abstract}

\section{Introduction}
Electromagnetic quasistatic (EMQS) field models are regarded as approximations to the full set of Maxwell equations if the structures under consideration are small w.r.t. to the wave lengths of the highest operational frequency for a given problem, i.e., if radiation effects can be neglected \cite{bHausMelcher:01s}.
The quasistatic field model regime includes static models, resulting in electrostatic and magnetostatic fields, where merely capacitive, resistive or inductive effects are considered. In addition, it also includes slowly varying electric or magnetic fields, such as those resulting from the electro-quasistatic field model, which covers problems governed by capacitive and resistive effects, and the magneto-quasistatic field model, which captures resistive and inductive effects. The majority of these established submodels of Maxwell's equations are based on mathematical formulations in terms of electric and/or magnetic scalar and vector potentials \cite{SteinmetzKurzClemens2011:01s}.
Quasistatic field problems, where resistive, inductive, and also capacitive effects need to be considered, have recently become of interest for high-frequency coil modeling, e.g., in transformers or in inductive wireless power transfer systems \cite{inpBadicsetal2018:01s}, and for studying electromagnetic compatibility aspects of power electronic systems. 
Although geometric multiscale aspect ratios of such systems are often in favor of lumped parameter circuit formulations, hybrid field-circuit formulations or circuit-type formulations, such as the partial-element equivalent circuit (PEEC) methods, and rather recently field oriented formulations have come into focus. 

Based on the original quasistatic field model of Darwin \cite{Darwin1920:01s}, \cite{RaviartSonnendruecker1995:01s2}, which was originally devised for describing the electromagnetic field of moving charged particles in free space without wave propagation effects, various generalizations of this field model capable of including conductive, dielectric and ferromagnetic material behavior based on scalar/vector potential formulations have been developed and analyzed in \cite{inpBadicsetal2018:01s}, \cite{KochWeiland2011:01s, KochSchneiderWeiland2012:01s, inpGarcia2018Chapter1:01s, ZhaoTang2019:01s,  BadicsPavoBiliczGyimothy2020:01s, inpClemensKaehneSchoeps2019:01s,  ClemensKasolisHenkelKaehneGuenther2021:01s, inpKaimori2020:01s, TahaTangHenneronLeMenachSalomezDucreux2021s}. 
For computational simulations of these quasistatic fields, where the conductive currents, the electric field in the non-conductive regions and the related magnetic field have to be described simultaneously, formulations that will result in symmetric algebraic systems of equations are favorable as they allow the use of numerical solution schemes that are more efficient than those for non-symmetric systems of equations. 

The paper is organized as follows. After this introduction, Darwin EMQS field models are presented, including discrete variants that are analyzed w.r.t. the symmetry of the resulting algebraic systems of equations.
Section three shows numerical experiments and comparisons of discrete symmetric formulations for the calculation of electromagnetic quasistatic fields in the Darwin model regime and is followed by a conclusion.

\section{Darwin-type EMQS Field Models}\label{sec:darwin}
Although the original Darwin model \cite{Darwin1920:01s} was derived for the description of quasistatic electromagnetic fields of plasmas, i.e., electric space charges in motion, in free space, more recent Darwin-type field models enable the consideration of non-homogeneous material distributions. The resulting EMQS formulations can be employed for modeling wireless charging systems, high-frequency coils, and power electronic systems, while they are useful for electromagnetic compatibility analysis of complex electric or electronic systems where radiation effects can be neglected.

Within Darwin-type EMQS field formulations the electric field $\bm{E}$ is commonly decomposed as 
\begin{equation}
\bm{E} = \bm{E}_{\mathrm{irr}} + \bm{E}_{\mathrm{rem}},
\label{eq:E_split}
\end{equation}
i.e., into an irrotational part  $\bm{E}_{\mathrm{irr}}= - \grad \varphi$ represented as a gradient of an electric scalar potential $\varphi$ and a remainder part  $\bm{E}_{\mathrm{rem}} = - \partial\bm{A}/\partial t$ represented via the time derivative of the magnetic vector potential $\bm{A}$ \cite{Larsson2007:01s}, which may also include those irrotational components of the electric field that are not represented by $\bm{E}_{\mathrm{irr}}$.
This is due to the fact, that some EMQS field formulations rely on the use of Coulomb(-type) gauges of the form $\divergence \left(\alpha (\bm{r}) \partial\bm{A}/\partial t \right) = 0,$ corresponding to an eventually locally varying material parameter $\alpha = \alpha (\bm{r})$, depending on the chosen field formulation. 

According to (\ref{eq:E_split}), the electromagnetic field is represented by
\begin{equation}
\bm{E} = - \frac{\partial}{\partial t} \bm{A} -\grad \varphi\text{,}\qquad
\bm{B} = \curl \bm{A}
\label{eq:E_and_B}
\end{equation}
where $\bm{E}$ denotes the electric field intensity and $\bm{B}$ the magnetic flux density. 

It is assumed that the field problem under consideration is defined in a bounded computational domain $\Omega$ that constitutes of a region $\Omega_c$ that contains electrically conductive materials with $\kappa > 0$, and a region $\Omega_n$ that contains electrically non-conductive material with $\kappa = 0$. While dielectric material parameters $\varepsilon$ are physically defined throughout $\Omega,$ some EMQS models restrict the consideration of the permittivity to the non-conductive regions $\Omega_n$ \cite{inpBadicsetal2018:01s}, \cite{BadicsPavoBiliczGyimothy2020:01s}. 

Equation \eqref{eq:E_and_B} together with the Amp\`{e}re law associated with the full set of Maxwell equations yields
\begin{equation}
\curl (\nu \curl \bm{A}) + \kappa \frac{\partial}{\partial t} \bm{A} 
                         + \kappa \grad \varphi
                         + \varepsilon \frac{\partial^2}{\partial t^2} \bm{A} 
                         + \varepsilon  \grad \frac{\partial}{\partial t} \varphi 
                        = \bm{J}_\mathrm{S},
\label{eqn_Ampere_A-phi}
\end{equation}
where $\nu=\mu^{-1}$ is the reluctivity, i.e., the inverse permeability, $\kappa$ denotes the electric conductivity, $\varepsilon$ the permittivity, and $\bm{J}_\mathrm{S}$ a source current density, respectively.

Applying the divergence to (\ref{eqn_Ampere_A-phi}) yields the full Maxwell continuity equation expressed as
\begin{equation}
\divergence \left(\kappa \frac{\partial}{\partial t} \bm{A} 
                        + \kappa \grad \varphi
                        + \varepsilon \frac{\partial ^2}{\partial t^2} \bm{A}
                        + \varepsilon  \grad \frac{\partial}{\partial t} \varphi\right)
                        = \divergence \bm{J}_\mathrm{S},
\label{eqn_fullMaxwell continuity}
\end{equation}
which is implicitly given by (\ref{eqn_Ampere_A-phi}), i.e., any solution pair  $(\bm{A}, \varphi)$ of the electrodynamical potentials that satisfies (\ref{eqn_Ampere_A-phi}) is also a solution to (\ref{eqn_fullMaxwell continuity}). Hence, the uniqueness of $(\bm{A}, \varphi)$ requires an additional gauge, either on the magnetic vector potential $\bm{A}$ or on the electro-quasistatic scalar potential $\varphi$, which can be imposed with suitable Dirichlet-type boundary conditions. Given the special case of a frequency domain reformulation chosen for time harmonic fields, uniqueness of the complex amplitude of $\bm{A}$ and $\varphi$ additionally requires that no resonance frequencies (i.e., frequencies of eigenmodes) are excited. In terms of numerical schemes this also includes the problem of so called low-frequency instabilities of related discrete field formulations corresponding to increasing condition numbers for angular frequencies $\omega$ tending to the static limit case, i.e., $\omega \rightarrow 0$.

Within the Darwin quasistatic electromagnetic field model \cite{Larsson2007:01s}, the rotational parts of the displacement current densities responsible for the modeling of radiation effects are neglected, i.e., $\varepsilon \partial^2 \bm{A}/\partial t^2 \cong \bm{0}$ . 
Under this assumption, (\ref{eqn_Ampere_A-phi}) reduces to the Darwin-Amp\`{e}re equation 
\begin{equation}
\curl (\nu \curl \bm{A}) + \kappa \frac{\partial}{\partial t} \bm{A} 
                        + \kappa \grad \varphi
                        + \varepsilon  \grad \frac{\partial}{\partial t} \varphi 
                        = \bm{J}_\mathrm{S},
\label{eqn_Darwin-Ampere}
\end{equation}
which features only first order time derivatives.
The definition in (\ref{eq:E_and_B}) requires additional gauging of both the magnetic vector potential and  the scalar electric potential. Special consideration is required for the electric field approximation in (\ref{eq:E_and_B}), as the irrotational parts of $\bm{A}$ are no longer regular.

The choice of possible different gauge conditions will result in distinguishable Darwin field models, i.e., Darwin field formulations are gauge dependent and may even describe different electric and magnetic field solutions as approximations to those described by the full set of Maxwell equations.

\subsection{The Original Darwin Field Model}\label{sec:orig_darwin}

The Darwin field model \cite{Darwin1920:01s} was introduced for modeling the electromagnetic quasistatic field of moving charges within electric plasmas in free space, i.e., with material parameters $\kappa=0$, $\varepsilon = \varepsilon_0$ and $\nu=\nu_0$. The formulation of this Darwin model in \cite{RaviartSonnendruecker1995:01s2} in terms of the electrodynamic potentials in \eqref{eq:E_and_B} relies on a full Helmholtz decomposition of the electric field intensity $\bm{E}$ into an irrotational part and a fully solenoidal part, i.e., assuming $\curl \bm{E}_{\mathrm{irr}}=0$ and $\divergence \bm{E}_{\mathrm{rem}}= - \divergence \left(\partial\bm{A} / \partial t \right)=0.$ The model additionally assumes a Coulomb-gauge $\divergence \bm{A}=0$ to hold and introduces the Darwin-Amp\`{e}re equation (\ref{eqn_Darwin-Ampere}) which features both $\bm{A}$ and $\varphi,$ i.e., an additional gauge equation is required for the solution.
As a consequence to the Helmholtz decomposition, the electric Gau{\ss} law chosen as second gauge equation does not consider the rotational parts of the electric field and consequently reduces to a Poisson equation. As a combined result, the Darwin electromagnetic quasistatic field formulation is given by the two equations
\begin{eqnarray}
-\nu_0 \Delta \bm{A} 
                        + \varepsilon_0  \grad \frac{\partial}{\partial t} \varphi 
                        &=& \rho_\mathrm{E} \bm{v},\\
        -\Delta \varphi 
                        &=& \rho_{\mathrm{E}} / \varepsilon_0,
\label{eqn_Original_Darwin}
\end{eqnarray}
%&
where the field sources are the electric charge density $\rho_{\mathrm{E}}$ and its current density vector $\bm{J}_\mathrm{S} = \rho_\mathrm{E} \bm{v},$ i.e., the motion of the charges along the direction given by a velocity vector $\bm{v}$.

\subsection{EMQS Fields and Full Maxwell Field Models}\label{sec:darwin_}

To eliminate the restriction of the original Darwin model (\ref{eqn_Original_Darwin}) to charge densities in motion as field sources, and in order to extend the EMQS model to include electrically conductive, dielectric and permeable materials, alternative gauging equations are considered, which result in different electromagnetic quasistatic Darwin-type approximations of the Maxwell equations, which are fully represented by the coupled system of partial differential equations featuring both Amp\`{e}re's equation (\ref{eqn_Ampere_A-phi}) and the full Maxwell continuity equation (\ref{eqn_fullMaxwell continuity}), respectively.

Alternatively, in \cite{OstrowskiHiptmair:2021s}, a full Maxwell electromagnetic field model based on (\ref{eqn_Ampere_A-phi}) including radiation effects is presented assuming an electro-quasistatic gauge which results in a two-step formulation, where  (\ref{eqn_fullMaxwell continuity}) is solved for the magnetic vector potential with known source current densities $\bm{J}_\mathrm{S}$ and the scalar potential $\varphi$ resulting from a preceeding solution of an electro-quasistatic source field problem.

This two-step frequency domain formulation \cite{OstrowskiHiptmair:2021s} also includes additional modifications of the model to address the low frequency instability of both the electro-quasistatic field model and the Amp\`{e}re curlcurl equation involved in this two-step full Maxwell system, respectively. A time domain formulation of this two-step full Maxwell field model is used with an implicit Newmark-beta time integration scheme in \cite{inpHenkelKasolisClemens2021:01s}. 
For time harmonic electromagnetic fields, another low-frequency stabilized full Maxwell formulation is presented in \cite{EllerReitzingerSchoepsZaglemayr2017:01s}. 

The established availability of the low-frequency stabilized full Maxwell field models in \cite{EllerReitzingerSchoepsZaglemayr2017:01s} and \cite{OstrowskiHiptmair:2021s} mostly reduces the practical impact of Darwin-type EMQS frequency domain field formulations in volume type discretization schemes to academic situations. Corresponding time domain EMQS formulations neglecting radiation effects, however, potentially avoid the stiffness of the full set of Maxwell equations that limits established time domain wave solvers such as the finite difference time domain (FDTD) scheme \cite{Yee1966:01s} while opening up possibilities to easily include localized nonlinear material characteristics of ferromagnetic or semi-conductive dielectric materials.

\subsection{The Darwin Continuity Equation}\label{subsec:darwin_continuity}

A modification of the original Darwin model to include dielectric, ferromagnetic and electrically conducting materials results from using a variant gauge equation. 
This is derived in \cite{KochWeiland2011:01s}, \cite{KochSchneiderWeiland2012:01s} with an application of the divergence operator to the Darwin-Amp\`{e}re equation (\ref{eqn_Darwin-Ampere}) that yields the Darwin continuity equation
\begin{equation}
\divergence \left(\kappa \frac{\partial}{\partial t} \bm{A} 
                        + \kappa \grad \varphi
                        + \varepsilon  \grad \frac{\partial}{\partial t} \varphi\right)
                        = \divergence \bm{J}_\mathrm{S}.
\label{eqn_Darwin-continuity}
\end{equation}
and results in a modified Darwin EMQS field formulation consisting of the coupled (\ref{eqn_Darwin-Ampere}) and (\ref{eqn_Darwin-continuity}).
The Darwin continuity equation lacks the expression $\divergence (\varepsilon \partial^2 \bm{A}/\partial t^2)$ which is present in the full Maxwell continuity equation.
Since the Darwin continuity equation is implicitly given by the Darwin-Amp\`{e}re equation (\ref{eqn_Darwin-Ampere}) from which it results after left-application of the divergence operator, any solution $(\bm{A},\varphi)$ of (\ref{eqn_Darwin-Ampere}) will also be a solution of (\ref{eqn_Darwin-continuity}), and thus, an additional gauge expression is required for the magnetic vector potential of the problem.

For this case, a gauge can be activated by assuming a small artificial electric conductivity value $\hat{\kappa}$ defined in the total computational domain $\Omega$ with $\kappa \gg \hat{\kappa}>0,$ in $\Omega,$ and adding the Coulomb-type gauge
\begin{equation}
\divergence \left( \hat{\kappa} \frac{\partial}{\partial t} \bm{A}\right)=0,
\label{eqn_Coulomb-type_gauge_kappa-hat}
\end{equation}
to (\ref{eqn_Darwin-continuity}) to yield the augmented Darwin continuity equation 
\begin{equation}
\divergence \left((\kappa+\hat{\kappa}) \frac{\partial}{\partial t} \bm{A} 
                        + \kappa \grad \varphi
                        + \varepsilon  \grad \frac{\partial}{\partial t} \varphi\right)
                        = \divergence \bm{J}_\mathrm{S}.
\label{eqn_Darwin-continuity-modified}
\end{equation}
The Coulomb-type gauge equation (\ref{eqn_Coulomb-type_gauge_kappa-hat}) is then only implicitly enforced by the difference of the Darwin continuity equation in the Darwin-Amp\`{e}re equation (\ref{eqn_Darwin-Ampere}) and the augmented Darwin continuity equation (\ref{eqn_Darwin-continuity-modified}) within the EMQS formulation of these two coupled equations.

Alternatively, $\hat{\kappa}$ can be added within the  Darwin-Amp\`{e}re equation \cite{KochWeiland2011:01s, KochSchneiderWeiland2012:01s} by 
\begin{equation}
\curl (\nu \curl \bm{A}) + (\kappa+\hat{\kappa}) \frac{\partial}{\partial t} \bm{A} 
                         + \kappa \grad \varphi
                         + \varepsilon  \grad \frac{\partial}{\partial t} \varphi 
                         = \bm{J}_\mathrm{S},
\label{eqn_Darwin-Ampere_regularized}
\end{equation}
which implies the extended Darwin continuity equation (\ref{eqn_Darwin-continuity-modified}). In the EMQS field formulations coupling (\ref{eqn_Darwin-Ampere_regularized}) and (\ref{eqn_Darwin-continuity}),  the gauge equation 
(\ref{eqn_Coulomb-type_gauge_kappa-hat}) is then again enforced by the difference of the implicitly and explicitly stated two  continuity equations as a regularization to the vector potential.

In both variants, the EMQS formulations consisting of the coupled systems of the Darwin-Amp\`{e}re equation (\ref{eqn_Darwin-Ampere}) and the extended Darwin continuity equation (\ref{eqn_Darwin-continuity-modified}) and also with the modified Darwin-Amp\`{e}re equation (\ref{eqn_Darwin-Ampere_regularized}) and Darwin continuity equation (\ref{eqn_Darwin-continuity}) 
are {\it implicitly} regularized, i.e., the Coulomb-type gauge (\ref{eqn_Coulomb-type_gauge_kappa-hat}) only occurs in the coupled system as the difference between the implicitly and the explicitly stated continuity equations. 
This approach of relying on an implicit enforcement of the Coulomb-type gauge requires the use of spatial discrete schemes where the discrete field formulations mimic these properties accordingly.
Another prerequisite is that the resulting systems of discrete equations are solved with high accuracy albeit the bad conditioning of the formulations and in addition, numerical round-off effects need to be considered.   

Spatial discretization schemes with these properties are e.g. available with the finite
integration technique (FIT)
\cite{Weiland1996:01s} or similar mimetic schemes as the finite
element method (FEM) using lowest-order N{\'e}d{\'e}lec elements \cite{Nedelec80:01s}, 
or the cell method (CM) \cite{Tonti2001:02s}.
Using the FIT notation, the coupled system of equations (\ref{eqn_Darwin-Ampere}) and (\ref{eqn_Darwin-continuity}) 
reformulates into the coupled systems of time continuous grid equations 
\begin{eqnarray}
   \!\!
   \C^{\trans} \fMnu   \C       \fitvec{a}
   +           \fMkap  \dndt    \fitvec{a}
   +           \fM_{\kappa}  \Gr     \bm{\upphi}
   +           \fMeps  \Gr   \dndt   \bm{\upphi}
   \!\!\!\!\!&=&\!\!\!\!\!\!
        \fitvec{j}_{\mathrm{s}},
        \label{FIT_Darwin1}
   \\
   \!\!
     \Gr^{\trans} \fMkap     \dndt  \fitvec{a}
   + \Gr^{\trans} \fMkap \Gr        \bm{\upphi}
   + \Gr^{\trans} \fMeps \Gr \dndt  \bm{\upphi}
   \!\!\!\!\!&=&\!\!\!\!\!\!
     \Gr^{\trans} \fitvec{j}_{\mathrm{s}},
        \label{FIT_Darwin2}
\end{eqnarray}
where $\fitvec{a}$ is the degrees of freedom (dof) vector related to
the line integrals of the magnetic vector potential along the grid edges, $\bm{\upphi}$ is the dof vector of
electric nodal scalar potentials, $\fitvec{j}_{\mathrm{s}}$ is a vector of transient source currents, $\C$ is the discrete curl operator
matrix, $\Gr$ and $\Gr^{\trans}$ are discrete gradient and
(negative) divergence operator matrices. The matrices $\fMnu$,
$\fMkap$, $\fMeps$ are discrete material matrices of reluctivities,
conductivities and permittivities, respectively, and the
construction of these discrete Hodge operators depends on the
specific discretization scheme.

Assuming a time harmonic field situation with an angular frequency $\omega,$  the frequency domain formulation of  the discrete Darwin-type EMQS equations (\ref{FIT_Darwin1}) and
(\ref{FIT_Darwin2}) with the definition of  
$\fM_{\sigma}:=\fMkap + j \omega \fMeps$ yields the singular algebraic system of equations
\begin{eqnarray}
   \left[
   \begin{array}{cc}
   \C^{\trans} \fMnu \C \!\! + \!\! j \omega \fM_{\kappa}    \!&\!              \fM_{\sigma} \Gr \!\!\\
   \Gr^{\trans} \fM_{\kappa}                                 \!&\! \frac{1}{j \omega } \Gr^{\trans} \fM_{\sigma} \Gr\!\!
   \end{array}
   \right]
  \left[
   \begin{array}{c}
   \!\fitvec{a}\!\\
   \!\bm{\upphi}\!
   \end{array}
   \right]
   \!\!=\!\!
  \left[
   \begin{array}{c}
   \!\fitvec{j}_{\mathrm{s}}\! \\
   \!\frac{1}{j \omega } \Gr^{\trans} \fitvec{j}_{\mathrm{s}}.\!
   \end{array}
   \right]
  % \nonumber\\
        \label{FIT_Darwin_Monolithic_FD}
\end{eqnarray}
The complex-valued system matrix in (\ref{FIT_Darwin_Monolithic_FD}) is singular, even in the case where the matrix of electrical conductivities $\fMkap$ is regular, as the Darwin continuity gauge 
(\ref{FIT_Darwin2}) is implicitly included in the discrete Darwin-Amp\`{e}re equation (\ref{FIT_Darwin1}) due to the property $\Gr^{\trans}\C^{\trans}=0$ of the discrete FIT incidence matrices \cite{KochWeiland2011:01s, KochSchneiderWeiland2012:01s}.

With the modification in the extended Darwin continuity equation (\ref{eqn_Darwin-continuity-modified}), the algebraic system (\ref{FIT_Darwin_Monolithic_FD}) is regularized by changing the left lower matrix block from $\Gr^{\trans}\fM_{\kappa}$ into $\Gr^{\trans}\fM_{\kappa+\hat{\kappa}}$, thereby enabling the use of direct algebraic system solvers. Its system matrix is still non-symmetric as 
$(\Gr^{\trans} \fM_{\kappa+\hat{\kappa}})^T \ne \fM_{\sigma} \Gr$ holds.

The block matrix in \eqref{eqn_Darwin-continuity-modified} can be extremely ill-conditioned
due to their off-diagonal matrix block entries varying by
many orders of magnitude. To improve the scaling, an equivalent reformulation with the scalar potential vector $\bm{\uppsi}$
with $\bm{\upphi} = j\omega \bm{\uppsi}$ corresponding to the representation of the vector of electric grid edge voltages $\fitvec{e} = -j\omega (\fitvec{a}+\Gr\bm{\uppsi} )$ yields the system
\begin{eqnarray}
   \left[
   \begin{array}{cc}
   \C^{\trans} \fMnu \C \!\! + \!\! j \omega \fM_{\kappa}    \!&\!  j\omega \fM_{\sigma} \Gr \!\!\\
   j \omega \Gr^{\trans} \fM_{\kappa+\hat{\kappa}}                                 \!&\!  j \omega \Gr^{\trans} \fM_{\sigma} \Gr\!\!
   \end{array}
   \right]
  \left[
   \begin{array}{c}
   \!\fitvec{a}\!\\
   \!\bm{\uppsi}\!
   \end{array}
   \right]
   \!\!=\!\!
  \left[
   \begin{array}{c}
   \!\fitvec{j}_{\mathrm{s}}\! \\
   \!\Gr^{\trans} \fitvec{j}_{\mathrm{s}}\!
   \end{array}
   \right]
%   \nonumber\\
        \label{FIT_Darwin_Monolithic_FD_regularized}
\end{eqnarray}
featuring an improved scaling of the matrix block entries while being mathematically equivalent to system \eqref{FIT_Darwin_Monolithic_FD}.

\subsection{Regularization and Symmetrization with Coulomb Gauge
Equations and the Darwin Continuity Equation}\label{subsec:timediscrete-full_continuity}

Alternative to the implicit regularization with a Coulomb-type gauge in (\ref{eqn_Coulomb-type_gauge_kappa-hat}) based on the introduction of a non-physical electrical conductivity parameter, a Coulomb-type gauge
\begin{equation}
\divergence \left(\varepsilon \frac{\partial}{\partial t} \bm{A}\right)=0,
\label{eqn_Coulomb-type_gauge}
\end{equation}
can be enforced by adding this term to the Darwin continuity equation \eqref{eqn_Darwin-continuity} after scaling it by a factor $\beta$ to achieve correct units. This approach used in both in \cite{ZhaoTang2019:01s} and \cite{inpKaimori2020:01s} yields the gauge equations
\begin{align}
\divergence \left(\kappa \frac{\partial}{\partial t} \bm{A} 
                        + \kappa \grad \varphi
                        + \varepsilon  \grad \frac{\partial}{\partial t} \varphi\right)&\phantom{=}\nonumber\\
                        +\beta\divergence \left(\varepsilon \frac{\partial}{\partial t} \bm{A}\right)
                        &= \divergence \bm{J}_\mathrm{S},
\label{eqn_semidiscrete-fullMaxwell continuity}
\end{align}
which can be interpreted as an approximation to the full Maxwell continuity equation (\ref{eqn_fullMaxwell continuity}) featuring only first order time derivatives.
The factor $\beta = \frac{\gamma}{\Delta t}$ is used both in \cite{ZhaoTang2019:01s} and \cite{inpKaimori2020:01s}, where $\gamma$ and $\Delta t$ are chosen already at this formulation level according to some subsequently used time integration scheme for the spatially discretized systems of equations. Following this approach, equation (\ref{eqn_semidiscrete-fullMaxwell continuity}) was labelled as a "semi-discrete" full Maxwell continuity equation \cite{inpClemensHenkelKasolisSchoeps2021:01s}.
Within one-step time marching schemes, the replacement of a second order time derivative expression 
$\varepsilon \partial^2\bm{A}/\partial t^2$ by $\varepsilon \frac{\gamma}{\Delta  t}\partial\bm{A}/\partial t$ does not affect the stability of the scheme, since the relation $\divergence (\varepsilon \frac{\gamma}{\Delta t}\bm{A}(t^{n+1}))=\divergence(\varepsilon \frac{\gamma}{\Delta t}\bm{A}(t^{n}))$ on the charge densities is maintained at each time step $t^{n+1},$ whereas a corresponding replacement of first order time derivatives in (\ref{eqn_semidiscrete-fullMaxwell continuity}) $(\partial/\partial t \rightarrow \gamma /\Delta t)$ eventually may result in instabilities during the time stepping process due to inconsistent charge accumulation \cite{inpClemensHenkelKasolisSchoeps2021:01s}. 

In addition, a suitable choice of the parameter $\beta$ according to the chosen implicit time discretization schemes results in symmetric system matrices of the form (with 
$\fM_{\sigma, \Delta t}:=\fMkap + \beta \fMeps$)
\begin{eqnarray}
    \left[
    \begin{array}{cc}
        \beta\fMkap+\C^{\trans} \fMnu   \C                                 & 
        \fM_{\sigma, \Delta t}\Gr\\
        \Gr^{\trans}  \fM_{\sigma, \Delta t}  &
        \frac{1}{\beta}  \Gr^{\trans} \fM_{\sigma, \Delta t}\Gr
    \end{array}
    \right]
    \left[
    \begin{array}{c}
          \fitvec{a}\\
         \bm{\upphi} 
    \end{array}
    \right]^{n+1}
    \!\!=\!\nonumber\\ 
    \operatorname{rhs}(\fitvec{a}^n,\bm{\upphi}^n,  \fitvec{j}_{\mathrm{s}}^{n+1},\Delta t).
    %\nonumber\\
    \label{FIT_monolithic_TD_symmetric}
\end{eqnarray}
In frequency domain formulations, the parameter choice $\beta = j \omega$ has the continuity equation (\ref{eqn_semidiscrete-fullMaxwell continuity}) coincide with the frequency domain version of the full Maxwell continuity equation (\ref{eqn_fullMaxwell continuity}) \cite{ZhaoTang2019:01s}, \cite{inpKaimori2020:01s} and the system matrix of discretized frequency domain formulation becomes symmetric with
\begin{eqnarray}
   \left[
   \begin{array}{cc}
   \C^{\trans} \fMnu \C \!\! + \!\! j \omega \fM_{\kappa}    \!&\!              \fM_{\sigma} \Gr \!\!\\
   \Gr^{\trans} \fM_{\sigma}                                 \!&\! \frac{1}{j \omega } \Gr^{\trans} \fM_{\sigma} \Gr\!\!
   \end{array}
   \right]
  \left[
   \begin{array}{c}
   \!\fitvec{a}\!\\
   \!\bm{\upphi}\!
   \end{array}
   \right]
   \!\!=\!\!
  \left[
   \begin{array}{c}
   \!\fitvec{j}_{\mathrm{s}}\! \\
   \!\frac{1}{j \omega } \Gr^{\trans} \fitvec{j}_{\mathrm{s}}\!
   \end{array}
   \right].
   %\nonumber\\
        \label{FIT_Darwin_Monolithic_FD_symmetric}
\end{eqnarray}
Similar to the Coulomb-type gauge \eqref{eqn_Coulomb-type_gauge_kappa-hat}, the Coulomb gauge (\ref{eqn_Coulomb-type_gauge}) only results in an implicit regularization within the coupled EMQS formulation featuring both the Darwin-Amp\`{e}re equation (\ref{eqn_Darwin-Ampere}) and 
the "semi-discrete" continuity equation in (\ref{eqn_semidiscrete-fullMaxwell continuity}), respectively.

In \cite[Section 6.3.1]{inpGarcia2018Chapter1:01s}, \cite{ZhaoTang2019:01s} and \cite{TahaTangHenneronLeMenachSalomezDucreux2021s}, the gauge equation (\ref{eqn_Coulomb-type_gauge}) is introduced and explicitly enforced using a Lagrange multiplier formulation. The corresponding discrete frequency domain formulation again features a complex-symmetric, non-Hermitean system matrix in   
\begin{eqnarray}
   \left[
   \begin{array}{ccc}
   \C^{\trans} \fMnu \C \!\! + \!\! j \omega \fM_{\kappa}    \!&\! \fM_{\sigma} \Gr 
                                                             \!&\! 
                                                                   j \omega \fM_{\varepsilon} \Gr\!\!\\
   \Gr^{\trans} \fM_{\sigma}                                 \!&\! 
   \frac{1}{j \omega } \Gr^{\trans}\fM_{\sigma} \Gr \!&\! \fitmat{0}\\ 
   j \omega\Gr^{\trans} \fM_{\varepsilon}           \!&\! \fitmat{0} \!&\! \fitmat{N} \!\!
   \end{array}
   \right]
  \left[
   \begin{array}{c}
   \!\fitvec{a}\!\\
   \!\bm{\upphi}\!\\
   \!\bm{\upgamma}\!
   \end{array}
   \right]
   \!\!=\!\!\nonumber\\
  \left[
   \begin{array}{c}
   \!\fitvec{j}_{\mathrm{s}}\! \\
   \!\frac{1}{j \omega } \Gr^{\trans} \fitvec{j}_{\mathrm{s}}\!\\
   \!\fitvec{0}\!
   \end{array}
   \right],
   %\nonumber\\
        \label{FIT_Darwin_Monolithic_FD_ShaoTang}
\end{eqnarray}
with $\fitmat{N}=0.$ As there exist solutions for $\fitvec{a}$ and $\bm{\upphi}$ with the nodal vector corresponding to the Lagrange multipiers $\bm{\upgamma}=0,$ the matrix $\fitmat{N}$ can be chosen as a positive definite regular matrix, i.e., $\fitmat{N}^{-1}$ is assumed to exist \cite{HaaseKuhnLanger2001s:01}. 
Applying a Schur-complement to eliminate the vector $\bm{\upgamma}$ yields 
\begin{eqnarray}
   \left[
   \begin{array}{cc}
   \fitmat{K} + j\omega \fM_{\kappa}   \!&\!              \fM_{\sigma} \Gr \!\!\\
   \Gr^{\trans} \fM_{\sigma}                                 \!&\! \frac{1}{j \omega } \Gr^{\trans} \fM_{\sigma} \Gr\!\!
   \end{array}
   \right]
  \left[
   \begin{array}{c}
   \!\fitvec{a}\!\\
   \!\bm{\upphi}\!
   \end{array}
   \right]
   \!\!=\!\!
  \left[
   \begin{array}{c}
   \!\fitvec{j}_{\mathrm{s}}\! \\
   \!\frac{1}{j \omega } \Gr^{\trans} \fitvec{j}_{\mathrm{s}}.\!
   \end{array}
   \right]
   %\nonumber\\
        \label{FIT_Darwin_Monolithic_FD_Grad-Div}
\end{eqnarray}
where the explicit Coulomb gauge now results in an additional discrete grad-div-operator to augment the left upper block matrix related to the discrete curlcurl operator with $\fitmat{K}=\!\!\C^{\trans} \fMnu \C \! + \!  
\omega^2  \fM_{\varepsilon} \Gr\fitmat{N}^{-1}\Gr^{\trans} \! \fM_{\varepsilon},$ similar to the regularized magneto-quasistatic curlcurl-equations in \cite{Bossavit2001:01s}, \cite{ClemensWeiland2002:02s}. 
The specific choice of the matrix $\fitmat{N}$ should consider that the effective spectral condition number of the matrix $\fitmat{K}$ is identical or close to that of the discrete curlcurl-matrix $\C^{\trans} \fMnu \C.$

\subsection{The Electro-Quasistatic Continuity Gauge Equation}\label{subsec:EQS_continuity}

In \cite{inpClemensKaehneSchoeps2019:01s, ClemensKasolisHenkelKaehneGuenther2021:01s}, the Darwin continuity equation (\ref{eqn_Darwin-continuity}) is reduced to the 
electro-quasistatic (EQS) continuity equation, 
\begin{equation}
\divergence \left( \kappa       \grad \varphi
                 + \varepsilon  \grad 
                   \frac{\partial}{\partial t} \varphi
                 \right)
                        = \divergence \bm{J}_\mathrm{S},
\label{eqn_EQS-continuity}
\end{equation} 
i.e., the EMQS model relies on the gauge of the electro-quasistatic potential similar the approach chosen for the full Maxwell formulation in \cite{OstrowskiHiptmair:2021s}.

The resulting Darwin-type EMQS formulation consisting of the Darwin-Amp\`{e}re equation \eqref{eqn_Darwin-Ampere} and \eqref{eqn_EQS-continuity} decouples and allows to formulate a two-step Darwin scheme. First and independently from the magnetic vector potential, the electro-quasistatic field problem (\ref{eqn_EQS-continuity}) is solved. 
The solenoidal total current densities $\bm{J}_\mathrm{EQS}=- \kappa       \grad \varphi - \varepsilon  \grad \frac{\partial}{\partial t} \varphi +\bm{J}_\mathrm{S}$ are then used as an excitation source to a (consistently) singular magneto-quasistatic (MQS) curlcurl formulation based on the magnetic vector potential $\bm{A}$ with
\begin{equation}
\curl (\nu \curl \bm{A}) + \kappa \frac{\partial}{\partial t} \bm{A} = \bm{J}_\mathrm{EQS},
\label{eqn_MQS-two-step-Ampere}
\end{equation}
where the irrotational components of $\bm{A}$ are initially not specified and require an additional regularization.
Whereas in the magneto-quasistatic case only the magnetic flux density is calculated, within EMQS field formulations the calculation of a electric field according to (\ref{eq:E_and_B}) requires a uniquely defined vector potential also in the non-conductive regions $\Omega_n.$ This requires an additional gauge for the magnetic vector potential in $\Omega_n$ \cite{ ClemensKasolisHenkelKaehneGuenther2021:01s}.

The combination of the Darwin-Amp\`{e}re equation (\ref{eqn_Darwin-Ampere}) and the implicitly included Darwin continuity equation (\ref{eqn_Darwin-continuity}) and the electro-quasistatic continuity gauge (\ref{eqn_EQS-continuity}) implicitly enforces the regularization of the magneto-quasistatic continuity equation 
\begin{equation}
\divergence \left(\kappa \frac{\partial}{\partial t} \bm{A}\right)=0,
\label{eqn_MQS-continuity}
\end{equation}
only acting in conductive regions $\Omega_c$ of the problem domain. The implicit enforcement of both the electro-quasistatic and the magneto-quasistatic continuity equation poses a limit to the modeling capabilities of the corresponding EMQS formulation \cite{inpOstrowskiWinkelmann:01s}. 
The formulation does not allow to model an inductively driven displacement of space charges to appear on the interface between  $\Omega_c$ and $\Omega_n,$ which act there as sources for irrotational electro-quasistatic field contributions in the non-conductive regions $\Omega_n$. The lack of continuity of normal components of the total current densities at conductor-/non-conductor interfaces, as they occur e.g. on capacitor surfaces within problem configurations, is a limit of the EMQS two-step scheme. In contrast, the full Maxwell two-step frequency domain formulation in \cite{OstrowskiHiptmair:2021s} models the corresponding irrotational electroquastatic field components in the second step using the full Maxwell-Amp\`{e}re equation featuring the expression $-\omega^2 \varepsilon \bm{A},$ i.e., in this formulation irrotational electro-quasistatic field parts are also represented in the magnetic vector potential. 

The introduction of an artificial conductivity value $\hat{\kappa}$ in $\Omega_n$ as suggested in \cite{ ClemensKasolisHenkelKaehneGuenther2021:01s} results in a regularisation of (\ref{eqn_MQS-two-step-Ampere}), and allows to consider approximations of the electric field according to (\ref{eq:E_and_B}) also with contributions of the vector potential $\bm{A},$ where the calculated amplitude vectors will include phase errors.  
The monolithic discrete frequency domain formulation of the EMQS model based on the EQS continuity equation (\ref{eqn_EQS-continuity}) given with
\begin{eqnarray}
   \!\!\left[
   \begin{array}{cc}
    \!\!\C^{\trans} \fMnu \C + j\omega \fM_{\kappa_{\Omega_c}+\hat{\kappa}_{\Omega_n}} 
                                              \!\!&\!\! \fM_{\sigma} \Gr \!\!\\
   \fitmat{0}                                 \!\!&\!\! \Gr^{\trans} \fM_{\sigma} \Gr\!\!
   \end{array}
   \right]\!
  \left[
   \begin{array}{c}
   \!\fitvec{a}\!\\
   \!\bm{\upphi}\!
   \end{array}
   \right]
   \!\!=\!\!
  \left[
   \begin{array}{c}
   \!\fitvec{j}_{\mathrm{s}}\! \\
   \! \Gr^{\trans} \fitvec{j}_{\mathrm{s}}\!
   \end{array}
   \right]\!\!
   %\nonumber\\
        \label{FIT_Darwin_Monolithic_FD_EQS-Gauge}
\end{eqnarray}
also includes the sequential frequency domain two-step Darwin EMQS scheme in terms of a block back-substitution scheme \cite{inpClemensHenkelKasolisSchoeps2021:01as}. 

The corresponding full Maxwell two-step formulation \cite{OstrowskiHiptmair:2021s} results from the exchange of 
$j\omega \fM_{\kappa_{\Omega_c}+\hat{\kappa}_{\Omega_n}}$ with $j\omega \fM_{\sigma}$ in (\ref{FIT_Darwin_Monolithic_FD_EQS-Gauge}),
\begin{eqnarray}
   \left[
   \begin{array}{cc}
    \C^{\trans} \fMnu \C + j\omega \fM_{\sigma}  
                                              \!\!&\!\! \fM_{\sigma} \Gr \!\!\\
   \fitmat{0}                                 \!\!&\!\! \Gr^{\trans} \fM_{\sigma} \Gr\!\!
   \end{array}
   \right]\!
  \left[
   \begin{array}{c}
   \!\fitvec{a}\!\\
   \!\bm{\upphi}\!
   \end{array}
   \right]
   \!\!=\!\!
  \left[
   \begin{array}{c}
   \!\fitvec{j}_{\mathrm{s}}\! \\
   \! \Gr^{\trans} \fitvec{j}_{\mathrm{s}}\!
   \end{array}
   \right]
   %\nonumber\\
        \label{FIT_FullMaxwell_Monolithic_FD_EQS-Gauge}
\end{eqnarray}
which is used to provide reference solutions for the numerical experiments in Section \ref{sec:numerical}.

\subsection{Domain Decomposition Continuity Gauge Equation }\label{subsec:DD_continuity}
The Darwin approximation in \cite{BadicsPavoBiliczGyimothy2020:01s} assumes different gauge equations in $\Omega_n$ and $\Omega_c$ (or $\Omega$), respectively. 
In the domain $\Omega_c$ related to conducting materials, merely a magneto-quasistatic continuity equation
\begin{equation}
\divergence \left(\kappa \frac{\partial}{\partial t} \bm{A} 
                 +\kappa \grad \varphi
            \right)
                        = 0
\label{eqn_MQS-continuity_eqn}
\end{equation} is assumed.
An additional gauge equation is Gauss's law 
\begin{equation}
\divergence \left(\varepsilon \frac{\partial}{\partial t} \bm{A} 
                 +\varepsilon \grad \varphi
            \right)
                        = -\rho_\mathrm{S}
\label{eqn_Gauss-continuity}
\end{equation}
considered just in the non-conductive regions $\Omega_c$ or, alternatively, in the total problem domain $\Omega$ \cite{inpClemensHenkelKasolisSchoeps2021:01s}. The excitation $\rho_\mathrm{S}$ is an electric charge density source term related to the excitation current densities via $\divergence \bm{J}_\mathrm{S} + \frac{\partial}{\partial t}\rho_\mathrm{S}=0$ and can be assumed to have zero value for many practical applications. 

An addition of (\ref{eqn_MQS-continuity_eqn}) and (\ref{eqn_Gauss-continuity}) assumed to hold in the total computational domain $\Omega$ includes a scaling factor $\beta = \gamma/\Delta t$ yields the combined gauge equation
\begin{equation}
\divergence \left( \left(\kappa \frac{\partial}{\partial t} \bm{A} 
                 + \kappa \grad \varphi\right)
                 \!+\! 
                  \beta
                  \left(
                  \varepsilon \frac{\partial}{\partial t} \bm{A}
                  +\varepsilon  \grad \varphi
                  \right)\right) 
                 =\beta \rho_\mathrm{S},
\label{eqn_DD-gauges-combined}
\end{equation}
which structurally resembles (\ref{eqn_fullMaxwell continuity}) and (\ref{eqn_semidiscrete-fullMaxwell continuity}). 
With a choice of a time integration scheme for both (\ref{eqn_Darwin-Ampere}) and (\ref{eqn_DD-gauges-combined}) corresponding to the parameter $\beta,$ a symmetric discrete Darwin EMQS formulations similar to  (\ref{FIT_monolithic_TD_symmetric}) and those given in \cite{inpKaimori2020:01s} can be formulated. 
If equation (\ref{eqn_Gauss-continuity}), however, is only defined in the non-conductive regions $\Omega_n$, e.g. by assuming negligible effects of displacement currents in conductive materials corresponding to the model in  \cite{BadicsPavoBiliczGyimothy2020:01s}, a symmetrization of the linear systems resulting from the spatially discrete formulation will additionally require a modification of the Darwin-Amp\`{e}re equation~\eqref{eqn_Darwin-Ampere}. 

\section{Numerical Experiments}\label{sec:numerical}
Two three-dimensional problems are considered --- a coil problem (Fig.~\ref{fig:geom1}) and a transformer problem (Fig.~\ref{fig:geom2}) with an open secondary circuit featuring a capacitor filled with a dielectric material. The material properties are summarized in Table~\ref{tab:dof}.
The computational domain $\Omega$ is free from charge and current sources. Both problems are driven by voltage excitation, that is, $\varphi_\mathrm{S}$ and $\varphi_\mathrm{G}$ are applied to the terminal surfaces $\Gamma_\mathrm{S}$ and $\Gamma_\mathrm{G}$, respectively. The terminal $\Gamma_\mathrm{G}$ is grounded and $\varphi_\mathrm{S}$ is $12~\mathrm{V}$ for the single coil problem and $0.1~\mathrm{V}$ for the transformer problem. The voltage excitation is imposed by Dirichlet boundary conditions, while homogeneous Neumann boundary conditions are prescribed on non-conducting surfaces. A homogeneous Dirichlet boundary condition $\bm{n}\times\bm{A}=\bm{0}$ is imposed for the magnetic vector potential on $\partial\Omega$.

\begin{figure}
\centering
\includegraphics[scale=0.78]{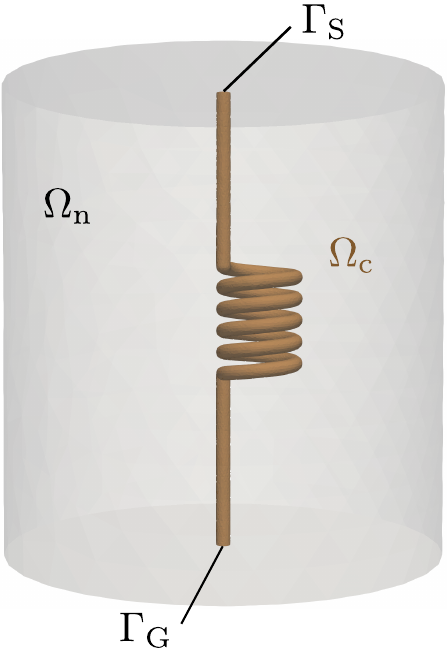}%\hfill
\caption{Computational domain of the coil problem. $\Omega_n$ is void and $\Omega_c$ is occupied by a conductor. $\Gamma_\mathrm{S}$ and $\Gamma_\mathrm{G}$ are the conductor terminal surfaces.}\label{fig:geom1}
\end{figure}
\begin{figure}
\centering
\includegraphics[scale=0.78]{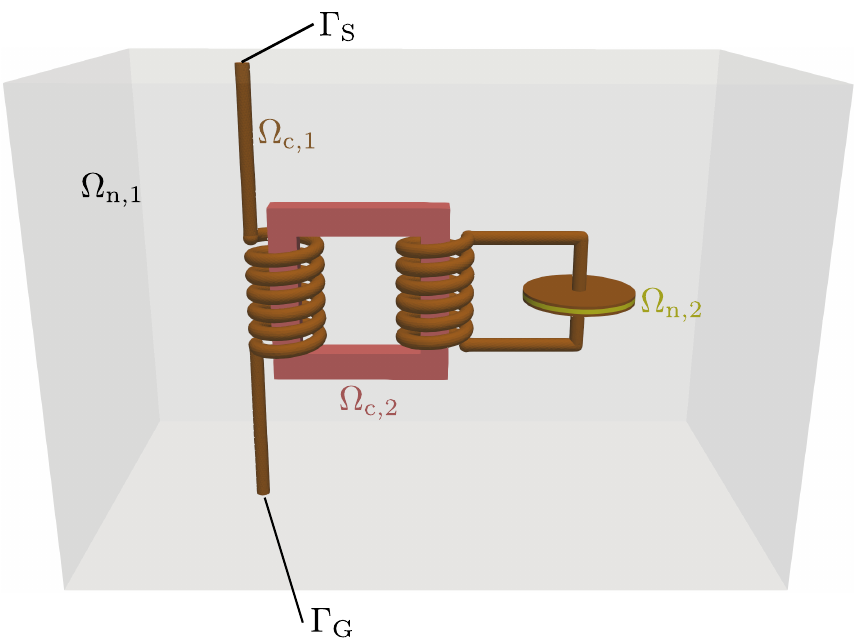}
\caption{Computational domain of the transformer problem, split into conductor $\Omega_{\mathrm{c},1}$, yoke $\Omega_{\mathrm{c},2}$, void $\Omega_{\mathrm{n},1}$, and dielectric $\Omega_{\mathrm{n},2}$. $\Gamma_\mathrm{S}$ and $\Gamma_\mathrm{G}$ are the conductor terminal surfaces.}
\label{fig:geom2}
\end{figure}
\begin{figure}
\centering
%[width=0.8\columnwidth]
\includegraphics{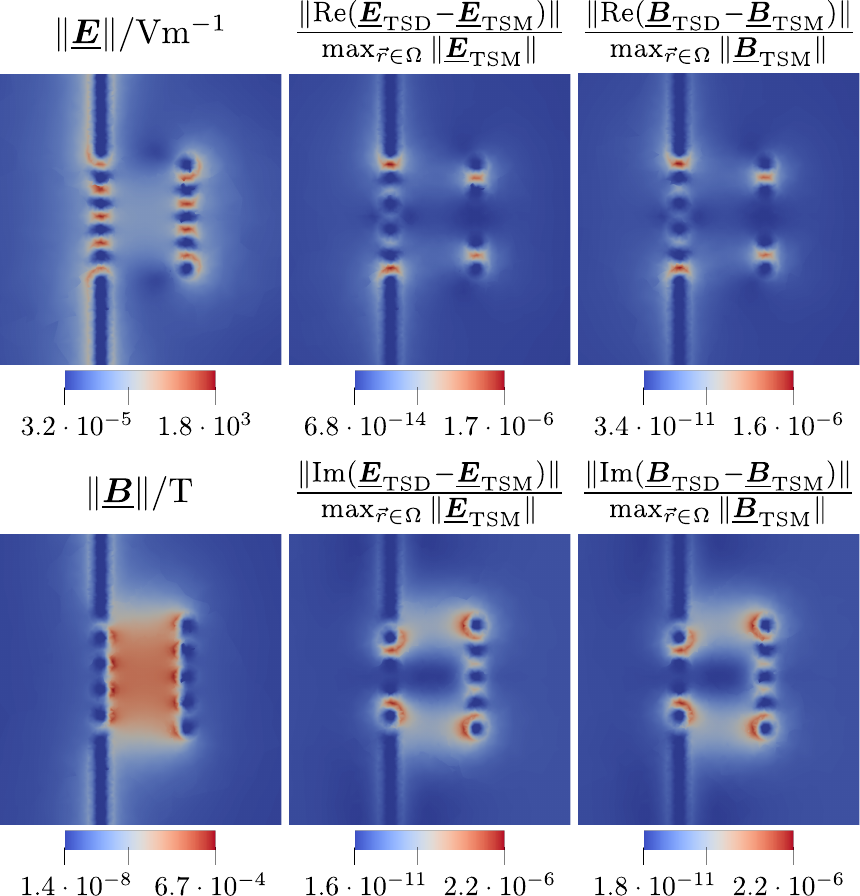}%\hfill
\caption{Left column: Magnitude of the electric field intensity and the magnetic flux density obtained by the two-step Darwin-type (TSD) EMQS method based on (\ref{eqn_EQS-continuity}) and (\ref{eqn_MQS-two-step-Ampere}). Middle and right column: Relative difference between two-step Darwin-type (TSD) EMQS scheme (based on (\ref{eqn_EQS-continuity}) and (\ref{eqn_MQS-two-step-Ampere})) and two-step full Maxwell (TSM) scheme (based on (\ref{eqn_Ampere_A-phi}) and (\ref{eqn_EQS-continuity})) for real and imaginary parts of electric field intensity and the magnetic flux density}\label{fig:helicalCoilTSDvsTSM}
\end{figure}
\begin{figure}
\centering
%[width=0.8\columnwidth]
\includegraphics{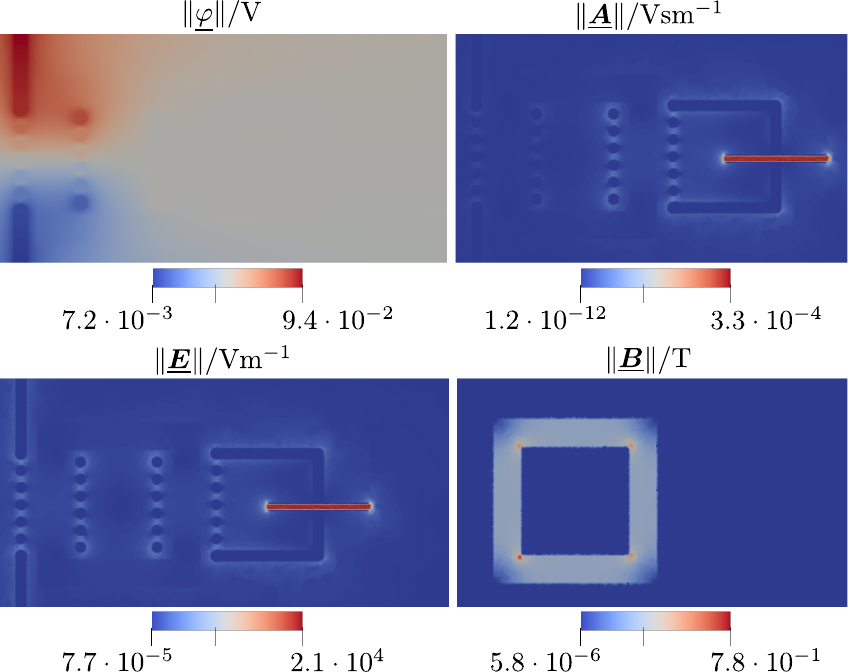}%\hfill
\caption{Magnitudes of electric scalar potential, magnetic vector potential, electric field intensity and magnetic flux density obtained by the two-step full Maxwell (TSM) formulation (based on (\ref{eqn_Ampere_A-phi}) and  (\ref{eqn_EQS-continuity})).}\label{fig:TrafoTSM}
\end{figure}
\begin{figure}
\centering
%[width=0.8\columnwidth]
\includegraphics{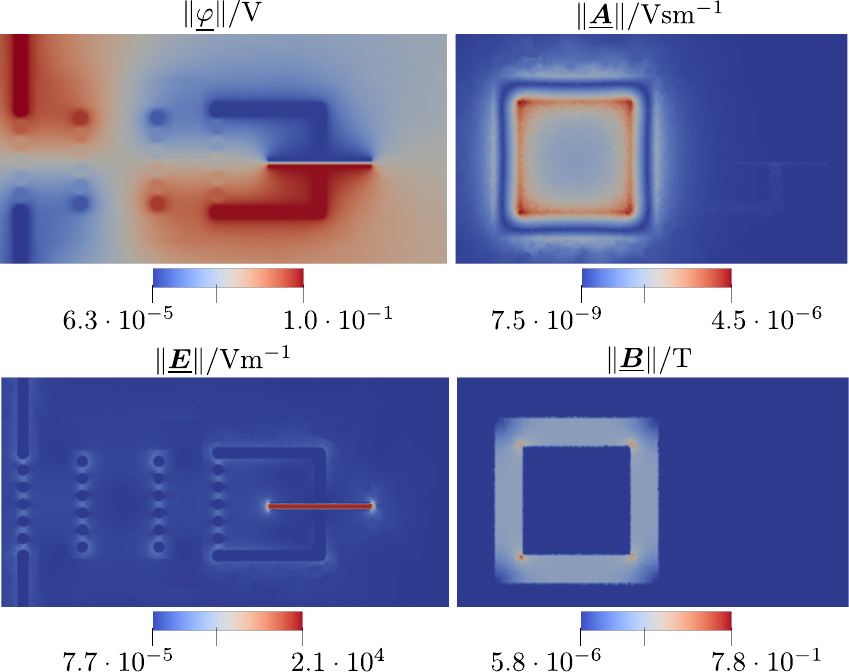}%\hfill
\caption{Magnitudes of electric scalar potential, magnetic vector potential, electric field intensity and magnetic flux density obtained by the full Maxwell continuity EMQS formulation based on (\ref{eqn_fullMaxwell continuity}) and (\ref{eqn_Darwin-Ampere}) discretized in (\ref{FIT_Darwin_Monolithic_FD_symmetric}) .}\label{fig:TrafoKWFullC}
\end{figure}
\begin{figure}
\centering
%[width=0.8\columnwidth]
\includegraphics{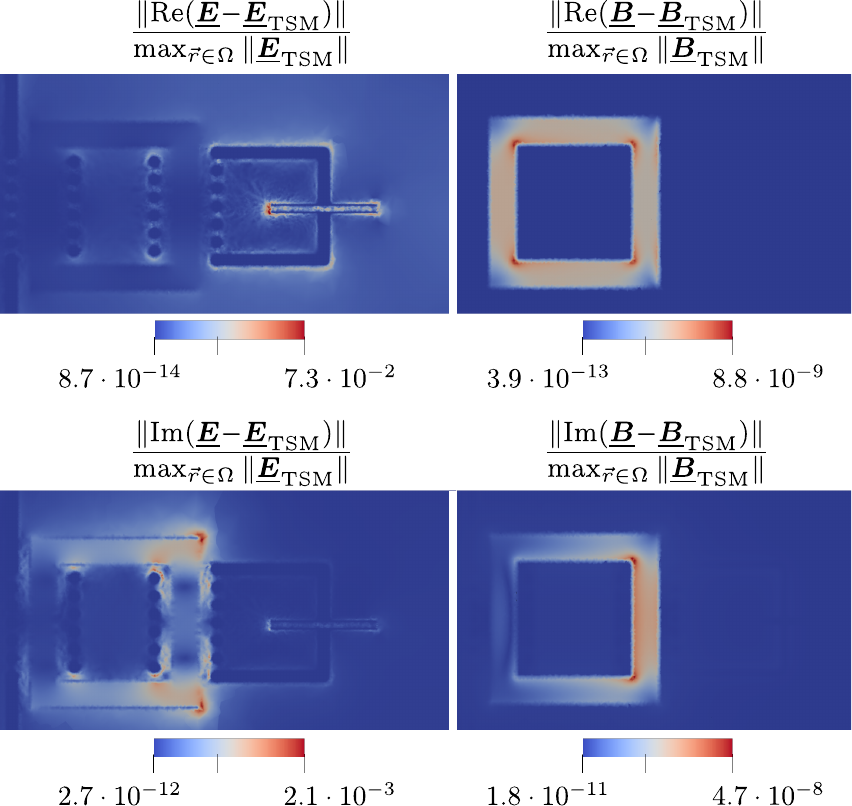}%\hfill
\caption{Relative difference between solutions of the reference two-step full Maxwell (TSM) formulation (combining (\ref{eqn_Ampere_A-phi}) and  (\ref{eqn_EQS-continuity})) and the full Maxwell continuity EMQS formulation (based on (\ref{eqn_fullMaxwell continuity}) and (\ref{eqn_Darwin-Ampere}) and discretized in  (\ref{FIT_Darwin_Monolithic_FD_symmetric})) for real and imaginary parts of the electric field intensity and the magnetic flux density .}\label{fig:TrafoPhiKWvsTS} 
\end{figure}

The problems are discretized using a mimetic finite element method with first-order Lagrange elements for the electric scalar potential and lowest-order N{\'e}d{\'e}lec elements for the magnetic vector potential. Table \ref{tab:dof} lists the number of degrees of freedom in each finite element space. All simulations are performed using in-house implementations in FreeFEM \cite{Hecht2012:01}, while the resulting linear systems are solved using the MUMPS direct solver library \cite{bMUMPS:01s}. Numerical experiments are conducted in frequency-domain for a frequency $f=10~\mathrm{MHz}$ with the corresponding wavelength in void being $\lambda \cong 30~\mathrm{m}$. The characteristic length of the helical coil is $6~\mathrm{cm}$ and that of the transformer is $20~\mathrm{cm}$. The reference solutions are obtained with the two-step full Maxwell (TSM) formulation \cite{OstrowskiHiptmair:2021s} implemented as a block-back-substitution of the system  (\ref{FIT_FullMaxwell_Monolithic_FD_EQS-Gauge}) using the same mesh. 

In the case of the coil problem, the physical fields obtained by the two-step Darwin-type (TSD) EMQS method based on (\ref{eqn_Darwin-Ampere}) and (\ref{eqn_EQS-continuity}) (Fig.~\ref{fig:TrafoTSM}) are compared with the reference fields (Fig.~\ref{fig:TrafoKWFullC}). The discrete TSD formulation reproduces the magnetic flux density of the coil as well as the parasitic capacitive effects between the coil windings with a maximum relative difference of $2.2\cdot10^{-6}$.

The transformer problem is investigated using the discrete frequency-domain full continuity EMQS formulation~\eqref{FIT_Darwin_Monolithic_FD_symmetric}. Since the reference formulation is based on the electro-quasistatic gauge which decouples the potentials $\bm{A}$ and $\varphi$, the electric scalar potential shown in Fig.~\ref{fig:TrafoTSM} yields only an electro-quasistatic current in the driving coil and the electric field between the capacitor plates is dominated by the contribution of the magnetic vector potential. 
In case of the full continuity EMQS  $(\bm{A}, \varphi )-$ formulation, the main contribution to the field between the capacitor plates stems from the scalar electric potential.
Both formulations yield approximately the same physical fields with a maximum relative difference for the real part of the electric field intensity of $7.3\cdot10^{-2}$ at the fringe of the capacitor and maximum relative difference for imaginary part of the magnetic flux density of $4.7\cdot10^{-8}$ at the inner corners of the yoke.

\begin{table}\centering
\caption{Material properties}
\label{tab:mat}
\begin{tabular}{lccc}
\toprule
&  $\sigma / \mathrm{Sm}^{-1}$ & $\mu_r$ & $\varepsilon_r$ \\
\midrule
Conductor & $5.96\cdot 10^{7}$ & $1$ & $1$\\
Yoke & $2\cdot 10^{-3}$ & $4000$ & $1$\\
Dielectric & $0$ & $1$ & $4$\\
Void & $0$ & $1$ & $1$\\
\bottomrule
\end{tabular}
\end{table}
\begin{table}\centering
\caption{Number of nodes (DOF) for each of the problems}
\label{tab:dof}
\begin{tabular}{lcc}
\toprule
& Lagrange elements & N{\'e}d{\'e}lec elements\\
\midrule
Coil & $40\,045$, & $285\,265$ \\
Transformer & $545\,546$ & $3\,924\,224$\\
\bottomrule
\end{tabular}
\end{table}

\section{Conclusions}
Electromagnetic quasistatic (EMQS) field model approximations for the full Maxwell field model allow to consider resistive, inductive and capacitive field effects, while ignoring radiation effects. Starting from the Darwin-Amp\`{e}re equation formulated in terms of the magnetic vector potential and electric scalar potentials $(\bm{A},\varphi)$, different EMQS field models were derived from variants of the Darwin continuity and the full Maxwell model continuity equation including implicitly and explicitly enforced Coulomb-type gauge expressions for the magnetic vector potential. The corresponding discrete field formulations expressed in terms of the mimetic Maxwell-grid-equations of the Finite Integration Technique 
were used to highlight the structure of the algebraic systems of equations resulting from frequency and time domain formulations. Numerical examples were presented to highlight the modeling capabilities of such EQMS field formulations in comparison to the full Maxwell model. 

\section{Acknowledgement}
 This work was supported in parts by the Deutsche Forschungsgemeinschaft (DFG) under grant no. CL143/11-2.

\footnotesize
\bibliographystyle{ieeetr}

\end{document}